\documentclass[11pt,twoside]{article}
\usepackage[pdftex]{graphicx}
\usepackage{amsmath}
\usepackage{amssymb}
\usepackage{cite}
\usepackage{epstopdf}

\usepackage{multicol}
\usepackage{amsmath}
\usepackage{relsize}
\usepackage[usenames]{color}

\begin{document}

\newcommand{\be}{\begin{equation}}
\newcommand{\ee}{\end{equation}}
\newcommand{\bm}{\boldmath}
\newcommand{\ds}{\displaystyle}
\newcommand{\bea}{\begin{eqnarray}}
\newcommand{\eea}{\end{eqnarray}}
\newcommand{\ba}{\begin{array}}
	\newcommand{\ea}{\end{array}}
\newcommand{\arcsinh}{\mathop{\rm arcsinh}\nolimits}
\newcommand{\arctanh}{\mathop{\rm arctanh}\nolimits}
\newcommand{\bc}{\begin{center}}
	\newcommand{\ec}{\end{center}}

\renewcommand{\labelenumi}{(\alph{enumi})} 
\let\vaccent=\v 
\renewcommand{\v}[1]{\ensuremath{\mathbf{#1}}} 
\newcommand{\gv}[1]{\ensuremath{\mbox{\boldmath$ #1 $}}} 
\newcommand{\uv}[1]{\ensuremath{\mathbf{\hat{#1}}}} 
\newcommand{\abs}[1]{\left| #1 \right|} 
\newcommand{\avg}[1]{\left< #1 \right>} 
\let\underdot=\d 
\renewcommand{\d}[2]{\frac{d #1}{d #2}} 
\newcommand{\dd}[2]{\frac{d^2 #1}{d #2^2}} 
\newcommand{\pd}[2]{\frac{\partial #1}{\partial #2}} 
\newcommand{\pdd}[2]{\frac{\partial^2 #1}{\partial #2^2}} 
\newcommand{\pdc}[3]{\left( \frac{\partial #1}{\partial #2}
	\right)_{#3}} 
\newcommand{\ket}[1]{\left| #1 \right>} 
\newcommand{\bra}[1]{\left< #1 \right|} 
\newcommand{\braket}[2]{\left< #1 \vphantom{#2} \right|
	\left. #2 \vphantom{#1} \right>} 
\newcommand{\matrixel}[3]{\left< #1 \vphantom{#2#3} \right|
	#2 \left| #3 \vphantom{#1#2} \right>} 
\newcommand{\grad}[1]{\gv{\nabla} #1} 
\let\divsymb=\div 
\renewcommand{\div}[1]{\gv{\nabla} \cdot #1} 
\newcommand{\curl}[1]{\gv{\nabla} \times #1} 
\let\baraccent=\= 
\renewcommand{\=}[1]{\stackrel{#1}{=}} 

	\begin{center} {\Large \bf
			\begin{tabular}{c}
				New information and entropic inequalities\\ for Clebsch-Gordan coefficients
			\end{tabular}			
		} 
		\begin{center}\textbf{
				V.~N.~Chernega, O.~V.~Manko, V.~I.~Manko, and Z. Seilov}
		\end{center}
		
		\end{center}
		
\begin{abstract}\noindent
			The Clebsch-Gordan coefficients of the group SU(2) are shown to satisfy new inequalities. They are obtained using the properties of Shannon and Tsallis entropies. The inequalities are obtained using the relation of squares of Clebsch-Gordan coefficients with probability distributions. Obtained inequalities are the new characteristics of correlations in quantum system of two spins.  The new inequalities were found for Hahn polynomials and hypergeometric functions.	
				
\end{abstract}

	
\begin{center}
	\section{Introduction}
\end{center}

The Clebsch-Gordan coefficients provide the possibility to solve the problem of obtaining the states of the system with angular momentum $j$ and the momentum projection $m=-j, -j+ 1, ... , j-1, j$ if the system is the composite system, containing two subsystems \cite{louck,LL,edmonds,vilenkin}. First subsystem is the system with angular momentum $j_1$ and momentum projection $m_1=-j_1, -j_1+ 1, ... , j_1-1, j_1$, the second subsystem is the system with angular momentum $j_2$ and momentum projection $m_2=-j_2, -j_2+ 1, ... , j_2-1, j_2$. From the group-theoretical point of view the Clebsch-Gordan coefficients give the solution of the problem related to the presenting in explicit form the product of two irreducible representations of the $SU(2)$-group as the sum of the irreducible representations of this group. The properties of the Clebsch-Gordan coefficients and 3-j symbols \cite{wigner} expressed in terms of the coefficients were intensively studied \cite{louck,suslov,smorodinsky,tolstoy}. The relation of Clebsch-Gordan coefficients and 3-j Wigner symbols to Hahn polynomials \cite{bateman,hahn} was found in \cite{karlin} in the 60s.

Recently it was shown \cite{chernega,manko1,manko2,manko3,manko4} that the the classical  probability distribution for single spin system (or single qudits) satisfies the information-entropic inequalities which have the form of subadditivity condition and the strong subadditivity condition (see, e.g., \cite{holevo}). These inequalities are known for composite systems containing subsystems. The subadditivity condition is the condition of nonnegativity of mutual information for bipartite composite systems. The strong subadditivity condition is the condition of nonnegativity of conditional information for tripartite systems.

The aim of this work is to obtain new inequalities for Clebsch-Gordan coefficients and 3-j symbols, which allow detecting new properties of these coefficients, not described in \cite{louck, suslov, smorodinsky, tolstoy}. These inequalities allow us to describe correlations in the system of two spins by means of mutual information in terms of Clebsch-Gordan coefficients. We note that physical content of Clebsch-Gordan coefficients and their physical properties can be clarified using the interpretation of these coefficients as the wave functions in image reconstruction procedure \cite{wolf}. We also obtain new inequalities  for Hahn polynomials using their relation to the Clebsch-Gordan coefficients \cite{nikiforov}. The inequalities are obtained due to known connection of squares of the coefficients with the probability distributions and applying the approach \cite{manko1} to information-entropic properties of the noncomposite system analogous to the approach known for composite system. In view of this analogy we apply the notions of mutual Shannon information, subadditivity property and Tsallis entropy to the probability distribution associated with Clebsch-Gordan coefficients. 

The paper is organized as follows:
In Sec.2 we review the properties of the Clebsch-Gordan coefficients and 3-j symbols. 
In Sec.3 we describe the information-entropic inequalities for bipartite systems and derive inequalities for Clebsch-Gordan coefficients. 
In Sec.4 we obtain new inequalities for Hahn polynomials.
In Sec.5 we give the conclusions and prospective. 

\section{Clebsch-Gordan coefficients}

The Clebsch-Gordan coefficients $\braket{j_1 m_1 j_2 m_2}{jm}$ are defined \cite{LL, edmonds} by the following relation :
\be 	\psi_{jm}=\sum_{m_1m_2}\braket{j_1 m_1 j_2 m_2}{jm} \psi_{j_1 m_1}^{(1)} \psi_{j_2 m_2}^{(2)} ,
\ee  where  $m= m_1+m_2$. Here $\psi_{jm}$ is the wave function of the spin system with spin $j$ and spin projection $m$; $\psi_{j_1 m_1}^{(1)}$ and $\psi_{j_2 m_2}^{(2)}$ are two wave functions of the spin system with spin $j_1$ and spin projection $m_1$, and the spin $j_2$ and spin projection $m_2$, respectively.
\\ The Wigner 3-j symbols $\begin{pmatrix}
j_1 & j_2 & j \\
m_1 & m_2 & -m 
\end{pmatrix}$ are defined as follows (see, e.g., \cite{LL}):
\be \braket{j_1 m_1 j_2 m_2}{jm} = (-1)^{j_1 - j_2 + m} \sqrt{2j + 1 } \begin{pmatrix}
	j_1 & j_2 & j \\
	m_1 & m_2 & -m 
\end{pmatrix}.
\ee

There are known properties of the 3-j symbols and Clebsch-Gordan coefficients reviewed, e.g., in \cite{louck,smorodinsky, asherova}.  For example, the 3-j symbols satisfies orthogonality relations
\be
\begin{aligned}
	 (2j + 1)  \mathlarger{\mathlarger{\sum}}\limits_{m_1 m_2}
\begin{pmatrix}
	j_1 & j_2 & j \\
	m_1 & m_2 & -m 
\end{pmatrix} \begin{pmatrix}
j_1 & j_2 & j' \\
m_1 & m_2 & -m' 
\end{pmatrix} = \delta_{jj'} \delta_{mm'},
\\
  \sum\limits_{j}
(2j + 1) \begin{pmatrix}
 	j_1 & j_2 & j \\
 	m_1 & m_2 & -m 
 \end{pmatrix} \begin{pmatrix}
 j_1 & j_2 & j \\
 m_1' & m_2' & -m 
\end{pmatrix} = \delta_{m_1 m_1'} \delta_{m_2 m_2'}.
\end{aligned}
\ee

These relations are expressed as orthogonality conditions for the Clebsch-Gordan coefficients 
\be
\begin{aligned} 
	\sum_{m_1 m_2} \braket{j_1 m_1 j_2 m_2}{jm} \braket{j_1 m_1 j_2 m_2}{j'm'} = \delta_{jj'} \delta_{mm'},
\\  \sum_{j} \braket{j_1 m_1 j_2 m_2}{jm} \braket{j_1 m_1' j_2 m_2'}{jm} = \delta_{m_1 m_1'} \delta_{m_2 m_2'}.
\end{aligned} 
\ee

 There is an explicit formula to 3-j symbols see, e.g., \cite{LL}:
 \begin{multline}\label{explicit}
 \begin{pmatrix}
 j_1 & j_2 & j_3 \\
 m_1 & m_2 & m_3 
 \end{pmatrix} = \left[ \frac{(j_1 + j_2 - j_3)! (j_1 - j_2+ j_3)! ((-j_1+j_2+j_3)!)}{(j_1+j_2+j_3+1)!}\right]^{1/2} \times \\
 \times [(j_1 + m_1)! (j_1 - m_1)! (j_2+m_2)!(j_2-m_2)!(j_3+m_3)!(j_3-m_3)!]^{(1/2)} \times \\
 \times \sum_{z}\left[ \left[ (-1)^{z+j_1-j_2-m_3}\right] \left[
 z!(j_1+j_2-j_3-z)!(j_1-m_1-z)! \times \right. \right.\\ \left.\kern-\nulldelimiterspace \left.\kern-\nulldelimiterspace
 \times (j_2+m_2-z)!(j_3-j_2+m_1+z)!(j_3-j_1-m_2+z)! \right]^{-1} \right].
 \end{multline}
 
 Here in the sum the numbers $z$ take all the integer values.
 
 Physical content of the Clebsch-Gordan coefficients is discussed, e.g., in review \cite{smorodinsky}. In papers \cite{jetp1997,dodonov} tomographic representation of spin states was introduced. In this representation spin states are described by probability distribution $\omega(m,\alpha, \beta)$ of spin projection $m=-j,-j+1,...,j-1,j$ on the direction given by a unitary vector $\vec{n}= (\sin{\beta}\cos{\alpha}, \sin{\beta}\sin{\alpha}, \cos{\beta})$. This probability distribution, called spin tomogram, defines the density matrix of quantum state $\rho_{mm'}$ and contains complete information about system state. The important role in tomographic approach in description of spin states belongs to the Clebsch-Gordan coefficients, because the explicit formulae showing connection of these coefficients to tomograms, can be checked experimentally.
 
 Namely, using 3-j symbols and Wigner D-functions we can find the relation between tomographic probability distribution $\omega(m_1,\alpha,\beta)$, defined as $$\omega(m_1,\alpha,\beta)=\sum_{m_1'=-j}^{j} \sum_{m_2'=-j}^{j} D_{m_1m_1'}^{(j)}(\alpha,\beta,\gamma=0) \rho_{m_1' m_2'}^{(j)} D_{m_1m_2'}^{(j)*}(\alpha,\beta,\gamma=0),$$ and density matrix of spin state $\rho_{m_1'm_2'}^{(j)}$ (see, e.g., \cite{jetp1997}):
 {\small
 \begin{multline}
 -\sum_{j_3=0}^{2j}\sum_{m_3=-j_3}^{j_3} (2j_3+1)^2 \sum_{m_1=-j}^{j} \int (-1)^{m_1} \omega(m_1, \alpha, \beta) D_{0m_3}^{(j_3)}(\alpha, \beta, \gamma=0) 
 \begin{pmatrix}
 j & j & j_3 \\
 m_1 & -m_1 & 0 
 \end{pmatrix}
 \begin{pmatrix}
 j & j & j_3 \\
 m_1' & -m_2' & m_3 
 \end{pmatrix} \frac{d \omega}{8 \pi^2} = \\ =(-1)^{m_2'} \rho_{m_1'm_2'}^{(j)}.
 \end{multline}}
 
 The relation between irreducible tensor operator $\hat{T}^{(j)}_{LM}$ of $SU(2)$ group and operator $\ket{jm}\bra{jm'}$ can also be expressed by means of Clebsch-Gordan coefficients (see, e.g., \cite{louck}):
 
 \be
 \hat{T}^{(j)}_{LM}=\sum_{m_1,m_2=-j}^{j} (-1)^{j-m_1}\braket{jm_2j(-m_1)}{LM} \ket{jm_2}\bra{jm_1},
 \ee
 \be
 \ket{jm}\bra{jm'}=\sum_{L=0}^{2j}\sum_{M=-L}^{L}  (-1)^{(j-m')} \braket{jmj(-m')}{LM}\hat{T}^{(j)}_{LM}.
 \ee
 
 The irreducible tensor operator in this form allows acquiring explicit form for the kernel of star-product tomographic symbols of physical observables for spin systems (see \cite{castanos}). The kernel of such star product is used to calculate statistical properties (such as mean values, higher moments, and correlations) of spin observables.
 
 For any selected $j_1$ and $j_2$ we can form matrix of Clebsch-Gordan coefficients, where columns corresponds to quantum numbers $j$ and $m$, and rows correspond to quantum numbers $m_1$ and $m_2$. The obtained matrix which we denote U is the unitary $N\times N$ matrix where $N=(2j_1 +1)(2j_2 +1)$. In fact, this matrix is orthogonal real matrix since the Clebsch-Gordan coefficients and 3-j symbols given by (\ref{explicit}) are real numbers. Any relations on Clebsch-Gordan coefficients both known \cite{louck,edmonds,wigner,smorodinsky} and obtained in this work help us to clarify the properties of measured statistical characteristics of spin states, determined by quantum tomograms.

\section{Inequalities for bipartite systems and Clebsch-Gordan coefficients}

	Matrix formed by squaring of each element of the unitary matrix $ U $ of the Clebsch-Gordan coefficients we denoted by $ \mathcal{B} $, i.e.: 
 \be \label{bistochastic} \bra{j_1 m_1 j_2 m_2} \mathcal{B} \ket{jm} \equiv |\braket{j_1 m_1 j_2 m_2}{jm}|^2 . \ee
 
 Let us make new notation for indices of matrix elements in the matrix ${\mathcal{B}}$:  $j_1  m_1 j_2 m_2 \leftrightarrow r$ and $jm \leftrightarrow s$. We label by the integers $ r = 1, 2, ... , N$, where $N=(2j_1 +1)(2j_2+1)$, the rows in the matrix $ \mathcal B $ shown by means of indices $ j_1  m_1 j_2 m_2 $ in Clebsch-Gordan coefficients, then we label by the same integers $ s=1, 2, ... ,N $ the columns in the matrix B, i.e., $ \bra{j_1 m_1 j_2 m_2} \mathcal{B} \ket{jm} \equiv \bra{r} \mathcal{B} \ket{s}$. We get then the matrix elements of the matrix $\mathcal{B}$ in usual form $\mathcal{B}_{rs}$, where both $ r $ and $ s=1, 2,..., N$. 
 
Obtained matrix is bistochastic matrix, which has the property that both columns and rows can be interpreted as probability distributions. It means that $ \mathcal{B}_{rs}\geq 0$ and normalization condition for elements in each rows and in each columns: 
 $ \sum_{r=1}^N \mathcal{B}_{rs} = \sum_{s=1}^N \mathcal{B}_{rs} =1 .$
  \\ We remind that for any discrete probability distribution $p_i$ $i=1,...,M$ the Shannon entropy is defined \cite{shannon}, and the entropy is nonnegative number which shows the rate of order in the system fluctuating observables: 
  \be \label{shannon}
  H(p_i)=-\sum_{i=1}^M p_i \log p_i.
  \ee
  
   In case of composite system its entropy satisfies variety of different properties. One of them is the subadditivity condition for Shannon entropy which has the following form: 
 \be \label{subadditivity} H(A) + H(B) \geq  H (AB) ,
 \ee where $ A $ and $ B $ are two subsystems of the composite system $ AB $. Probability distributions of these subsystems can be found as marginal probability distributions of composite system $AB$. Dimensions $M$ of subsystems $ A $, $ B $ and system $ AB $ in case of spin observables are $M=n_1=2j_1+1$,  $M=n_2=2j_2+1$ and $M=n_1 n_2$ respectively.
 \\Entropies of these systems are calculated using formula (\ref{shannon}).
 
 The Shannon entropy can be expressed in terms of Clebsch-Gordan coefficients for any row or column of matrix $\mathcal{B}$. 
  Condition (\ref{subadditivity}) can be represented in the following form:
  
 	\begin{multline}
 -\sum_{m_2=-j_2}^{j_2}|\braket{j_1 m_1 j_2 m_2}{jm}|^2 \log \left[ \sum_{m_2=-j_2}^{j_2}|\braket{j_1 m_1 j_2 m_2}{jm}|^2 \right] - \\ - \sum_{m_1=-j_1}^{j_1}|\braket{j_1 m_1 j_2 m_2}{jm}|^2 \log \left[ \sum_{m_1=-j_1}^{j_1}|\braket{j_1 m_1 j_2 m_2}{jm}|^2\right] \geq \\ \geq - \sum_{m_1=-j_1}^{j_1} \sum_{m_2=-j_2}^{j_2}|\braket{j_1 m_1 j_2 m_2}{jm}|^2 \log \left[\sum_{m_1=-j_1}^{j_1} \sum_{m_2=-j_2}^{j_2}|\braket{j_1 m_1 j_2 m_2}{jm}|^2 \right].
	\end{multline}

This inequality is the new characteristic of the Clebsch-Gordan coefficients, complementing known in literature \cite{louck,suslov,smorodinsky,tolstoy}. The notion of mutual information $I$ which is connected to the subadditivity condition is defined as follows:

\be \label{ish}
	 I = H(A) + H(B) - H (AB) \geq 0  .
\ee
The value of mutual information shows the rate of correlations of composite system with two subsystems. Using the expression (\ref{ish}) we introduce the concept of mutual information $I$ in terms the Clebsch-Gordan coefficients:
\begin{multline} \label{info}
I= \sum_{m_1=-j_1}^{j_1} \sum_{m_2=-j_2}^{j_2}|\braket{j_1 m_1 j_2 m_2}{jm}|^2 \log \left[\sum_{m_1=-j_1}^{j_1} \sum_{m_2=-j_2}^{j_2}|\braket{j_1 m_1 j_2 m_2}{jm}|^2 \right] - \\	-\sum_{m_2=-j_2}^{j_2}|\braket{j_1 m_1 j_2 m_2}{jm}|^2 \log \left[ \sum_{m_2=-j_2}^{j_2}|\braket{j_1 m_1 j_2 m_2}{jm}|^2 \right] - \\ - \sum_{m_1=-j_1}^{j_1}|\braket{j_1 m_1 j_2 m_2}{jm}|^2 \log \left[ \sum_{m_1=-j_1}^{j_1}|\braket{j_1 m_1 j_2 m_2}{jm}|^2\right]  \geq 0.
\end{multline}

In accordance with the main properties of mutual information in case of its vanishing there are no correlations in the system. For spin systems the rate of the correlation, described by the Clebsch-Gordan coefficients in the (\ref{info}) is defined by the value of mutual information $I$. Thus the physical content of acquired information-entropic inequalities describes presence and the rate of quantum correlations in the system of two spins.

\noindent For bipartite system also known is the Araki-Lieb inequality \cite{araki}:
\be \label{araki}
H (AB) \geq |H(A) - H(B)|	.
\ee

\noindent This inequality can be expressed in terms of Clebsch-Gordan coefficients:

\begin{multline}
	- \sum_{m_1=-j_1}^{j_1} \sum_{m_2=-j_2}^{j_2}|\braket{j_1 m_1 j_2 m_2}{jm}|^2 \log \left[\sum_{m_1=-j_1}^{j_1} \sum_{m_2=-j_2}^{j_2}|\braket{j_1 m_1 j_2 m_2}{jm}|^2 \right]\geq 
	\\  \geq \left|-\sum_{m_2=-j_2}^{j_2}|\braket{j_1 m_1 j_2 m_2}{jm}|^2 \log \left[ \sum_{m_2=-j_2}^{j_2}|\braket{j_1 m_1 j_2 m_2}{jm}|^2 \right] + \right.
	\\ \left.\kern-\nulldelimiterspace + \sum_{m_1=-j_1}^{j_1}|\braket{j_1 m_1 j_2 m_2}{jm}|^2 \log \left[ \sum_{m_1=-j_1}^{j_1}|\braket{j_1 m_1 j_2 m_2}{jm}|^2\right]  \right|.
\end{multline}

\noindent 
This inequality is the new entropic characteristics of the Clebsch-Gordan coefficients. One of the generalizations of Shannon entropy is the Tsallis entropy \cite{tsallis}, which is defined for probability distribution as follows:

\be
	T_q(p_i)=\frac{1}{1-q}(\sum_{i=1}^{M}p_i^q - 1),
\ee where q is called entropic index. If $ q\rightarrow 1$ the Tsallis entropy tends to the Shannon entropy.
\\For bipartite system the Tsallis entropy has the subadditivity property which reads:
\be \label{tssa}
	T_q(A,B) \leq T_q(A) + T_q(B).
\ee

\noindent 
For system of two spins the subadditivity condition of Tsallis entropy is expressed in terms of Clebsch-Gordan coefficients. Thus we obtain new inequality which is the new characteristics of Clebsch-Gordan coefficients, complementing known in \cite{louck,suslov,smorodinsky,tolstoy}:

\begin{multline}
	\frac{1}{1-q} \left[ \sum_{m_1=-j_1}^{j_1}\sum_{m_2=-j_2}^{j_2}|\braket{j_1 m_1 j_2 m_2 j_3 m_3}{jm}|^{2q} - 1 \right] \leq
	\\ \leq \frac{1}{1-q} \left[ \sum_{m_1=-j_1}^{j_1}|\braket{j_1 m_1 j_2 m_2 j_3 m_3}{jm}|^{2q} - 1 \right] + \frac{1}{1-q} \left[\sum_{m_2=-j_2}^{j_2}|\braket{j_1 m_1 j_2 m_2 j_3 m_3}{jm}|^{2q} - 1 \right].
\end{multline}

 In the Fig. 1 We show the dependence of Tsallis information $I_q= T_q(A) + T_q(B) -T_q(AB)$ on the entropic index $q$ for the case of two particles $j_1=5/2, ~~ j_2=2$ using the state $(j=9/2,m=1/2)$. 
\\ As for mutual Shannon information, Tsallis information takes nonnegative values. In this case for $q \rightarrow 1$ Tsallis information coincides with mutual information determined by Shannon entropy: $I_q \rightarrow I = 1.176$ in considered case. Obtained value, corresponding to mutual information is half of the value of its maximal value for considered case  $I_{max}=min\{H(A),H(B)\}=\log_2(5)=2.32$, but it is not equal to zero, and this shows the presence of correlations in the system of two spins. 

\begin{figure}[h]
	\noindent\centering{
		\includegraphics[width=70mm]{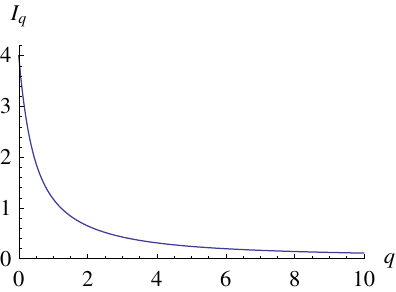}
	}
	\caption{The dependence of Tsallis information $I_q$ on the entropic index $q$ for selected $j_1=5/2,j_2=2,j=9/2,m=1/2$.}
	\label{pic1}
\end{figure}

\newpage
\section{Inequalities for Hahn polynomials}

The Hahn polynomial $h_{n}^{(\alpha \beta)} (x,N)$ may be defined in terms of generalized hypergeometric series\\ $_3 F_2 (a_1, a_2, a_3; b_1, b_2; z) = \sum_{k=0}^{\infty}\frac{(a_1)_k (a_2)_k(a_3)_k}{(b_1)_k (b_2)_k} \frac{z^k}{k!}$, where $(x)_k= \frac{\Gamma(x+k)}{\Gamma(x)}=x(x+1)...(x+k-1)$ is the Pochammer symbol. The Hahn polynomial reads:
\be \label{hahn}
	h_{n}^{(\alpha \beta)} (x,N) = \frac{(-1)^n(N-n)_n(\beta+1)_n}{n!} {_3 F_2}(-n, -x, \alpha+\beta+n+1;  \beta +1, 1-N; 1).
\ee
 
	The Clebsch-Gordan coefficients in terms of the Hahn polynomials may be defined as follows\cite{nikiforov, atakishiyev}:
\be \label{represent}
	(-1)^{j_1-m_1}\braket{j_1 m_1 j_2 m_2}{j m} = \frac{\sqrt{\rho(x)}}{d_n} h_n^{(\alpha \beta)}(x, N)= \frac{\sqrt{\rho(j_2-m_2)}}{d_{j-m}} h_{j-m}^{(m-j_1 +j_2,m+j_1 -j_2)}(j_2-m_2, j_1 +j_2 -m +1),
\ee
where 
$n = j-m$;
$x=j_2-m_2$; 
$N=j_1 +j_2 -m +1$; 
$\alpha=m-j_1 +j_2$; 
$\beta =m+j_1 -j_2$. \\
We used weight function
 $$ \rho(x)=\frac{\Gamma(N+\alpha-x)\Gamma(\beta + 1 +x)}{\Gamma(x+1)\Gamma(N-x)},~~~~~~~~~ \alpha> -1,~~~~~~~~~ \beta > -1 $$ 
 and squared norm $$d_n^2 = \frac{\Gamma(\alpha + n + 1) \Gamma(\beta + n+1) \Gamma(\alpha+\beta + n + N + 1)}{(\alpha+\beta+2n+1) n! (N-n-1)!\Gamma(\alpha+\beta + n+ 1)} .$$

\noindent Squared Clebsch-Gordan coefficients expressed in terms of Hahn polynomials reads:	$$|\braket{j_1 m_1 j_2 m_2}{jm}|^2= \rho(j_2-m_2)d_{j-m}^{-2} [h_{j-m}^{(m-j_1 +j_2,m+j_1 -j_2)}(j_2-m_2, j_1 +j_2 -m +1)]^2 .$$

	Using this representation of Clebsch-Gordan coefficients, we can rewrite  subadditivity inequality (\ref{subadditivity}) in terms of Hahn polynomials:

\begin{multline}
  	-\sum_{m_2=-j_2}^{j_2}\rho(j_2-m_2)d_n^{-2} [h_{j-m}^{(m-j_1 +j_2,m+j_1 -j_2)}(j_2-m_2, j_1 +j_2 -m +1)]^2 \times 
 	\\ \times  \log \left[ \sum_{m_2=-j_2}^{j_2}\rho(j_2-m_2)d_n^{-2} [h_{j-m}^{(m-j_1 +j_2,m+j_1 -j_2)}(j_2-m_2, j_1 +j_2 -m +1)]^2 \right] - 
 	\\ - \sum_{m_1=-j_1}^{j_1}\rho(j_2-m_2)d_n^{-2} [h_{j-m}^{(m-j_1 +j_2,m+j_1 -j_2)}(j_2-m_2, j_1 +j_2 -m +1)]^2 \times
 	 \\ \times \log \left[ \sum_{m_1=-j_1}^{j_1}\rho(j_2-m_2)d_n^{-2} [h_{j-m}^{(m-j_1 +j_2,m+j_1 -j_2)}(j_2-m_2, j_1 +j_2 -m +1)]^2\right] \geq 
 	 \\ \geq - \sum_{m_1=-j_1}^{j_1} \sum_{m_2=-j_2}^{j_2}\rho(j_2-m_2)d_n^{-2} [h_{j-m}^{(m-j_1 +j_2,m+j_1 -j_2)}(j_2-m_2, j_1 +j_2 -m +1)]^2 \times 
 	 \\\times \log \left[\sum_{m_1=-j_1}^{j_1} \sum_{m_2=-j_2}^{j_2}\rho(j_2-m_2)d_n^{-2} [h_{j-m}^{(m-j_1 +j_2,m+j_1 -j_2)}(j_2-m_2, j_1 +j_2 -m +1)]^2 \right].
 \end{multline}

Using the representation of Clebsch-Gordan coefficients in terms of Hahn polynomials (\ref{represent}) and applying the Araki-Lieb inequality (\ref{araki}) we get new relation for the Hahn polynomials in the following form:

\begin{multline}
	\sum_{m_1=-j_1}^{j_1} \sum_{m_2=-j_2}^{j_2}\rho(j_2-m_2)d_n^{-2} [h_{j-m}^{(m-j_1 +j_2,m+j_1 -j_2)}(j_2-m_2, j_1 +j_2 -m +1)]^2 \times 
	\\\times \log \left[\sum_{m_1=-j_1}^{j_1} \sum_{m_2=-j_2}^{j_2}\rho(j_2-m_2)d_n^{-2} [h_{j-m}^{(m-j_1 +j_2,m+j_1 -j_2)}(j_2-m_2, j_1 +j_2 -m +1)]^2 \right] \geq 
	\\  \geq  \left| \sum_{m_2=-j_2}^{j_2}\rho(j_2-m_2)d_n^{-2} [h_{j-m}^{(m-j_1 +j_2,m+j_1 -j_2)}(j_2-m_2, j_1 +j_2 -m +1)]^2 \times \right.
	\\ \times  \log \left[ \sum_{m_2=-j_2}^{j_2}\rho(j_2-m_2)d_n^{-2} [h_{j-m}^{(m-j_1 +j_2,m+j_1 -j_2)}(j_2-m_2, j_1 +j_2 -m +1)]^2 \right] - 
	\\ - \sum_{m_1=-j_1}^{j_1}\rho(j_2-m_2)d_n^{-2} [h_{j-m}^{(m-j_1 +j_2,m+j_1 -j_2)}(j_2-m_2, j_1 +j_2 -m +1)]^2 \times
	\\ \left.\kern-\nulldelimiterspace \times \log \left[ \sum_{m_1=-j_1}^{j_1}\rho(j_2-m_2)d_n^{-2} [h_{j-m}^{(m-j_1 +j_2,m+j_1 -j_2)}(j_2-m_2, j_1 +j_2 -m +1)]^2\right] \right|.
\end{multline}

	The subadditivity condition for Tsallis entropy (\ref{tssa}) can also be expressed in terms of the Hahn polynomials:

\begin{multline}
		\frac{1}{1-q} \left[ \sum_{m_1=-j_1}^{j_1}\sum_{m_2=-j_2}^{j_2}\rho(j_2-m_2)d_n^{-2} [h_{j-m}^{(m-j_1 +j_2,m+j_1 -j_2)}(j_2-m_2, j_1 +j_2 -m +1)]^{2q} - 1 \right] \leq
		\\ \leq \frac{1}{1-q} \left[ \sum_{m_1=-j_1}^{j_1}\rho(j_2-m_2)d_n^{-2} [h_{j-m}^{(m-j_1 +j_2,m+j_1 -j_2)}(j_2-m_2, j_1 +j_2 -m +1)]^{2q} - 1 \right] +
		\\+ \frac{1}{1-q} \left[\sum_{m_2=-j_2}^{j_2}\rho(j_2-m_2)d_n^{-2} [h_{j-m}^{(m-j_1 +j_2,m+j_1 -j_2)}(j_2-m_2, j_1 +j_2 -m +1)]^{2q} - 1 \right].
\end{multline}
	 
 Finally we obtain the inequality for hypergeometric function $_3F_2$.
Inequality (\ref{subadditivity}) rewritten in terms of hypergeometric functions reads: 

{ \small	\begin{multline}
 	-\sum_{m_2=-j_2}^{j_2} \left[ \left[\frac{(j_1+j_2-j)! (2j+1) \Gamma(j+j_1-j_2 +1)   \Gamma(m+j+1)\Gamma(j_2+m_2+1) \Gamma(j_1+m_1+1)}{(j-m)! \Gamma(j-j_1+j_2+1) \Gamma(j+j_1+j_2+2)\Gamma(j_2-m_2+1) \Gamma(j_1-m_1+1)} 
 	\right] \times \right.
 	\\ \left.\kern-\nulldelimiterspace \times \left[ \frac{\Gamma(j_1+j_2-m+1)}{\Gamma(-j+j_1+j_2+1)} \frac{{_3F_2({m-j},{m_2-j_2},{m+j+1};{m+j_1-j_2+1},{m-j_1-j_2};1 )}}{\Gamma(m+j_1-j_2+1)} \right]^2 \right]\times
 	\\ \times \log \left[ \sum_{m_2=-j_2}^{j_2}\left[ \left[\frac{(j_1+j_2-j)! (2j+1) \Gamma(j+j_1-j_2 +1)   \Gamma(m+j+1)\Gamma(j_2+m_2+1) \Gamma(j_1+m_1+1)}{(j-m)! \Gamma(j-j_1+j_2+1) \Gamma(j+j_1+j_2+2)\Gamma(j_2-m_2+1) \Gamma(j_1-m_1+1)}	
 	\right] \times \right.\right.
 	\\ \left.\kern-\nulldelimiterspace\left.\kern-\nulldelimiterspace \times \left[ \frac{\Gamma(j_1+j_2-m+1)}{\Gamma(-j+j_1+j_2+1)} \frac{{_3F_2({m-j},{m_2-j_2},{m+j+1};{m+j_1-j_2+1},{m-j_1-j_2};1 )}}{\Gamma(m+j_1-j_2+1)} \right]^2 \right]\right] - 
 	\\ - \sum_{m_1=-j_1}^{j_1}\left[ \left[\frac{(j_1+j_2-j)! (2j+1) \Gamma(j+j_1-j_2 +1)   \Gamma(m+j+1)\Gamma(j_2+m_2+1) \Gamma(j_1+m_1+1)}{(j-m)! \Gamma(j-j_1+j_2+1) \Gamma(j+j_1+j_2+2)\Gamma(j_2-m_2+1) \Gamma(j_1-m_1+1)}	
 	\right] \times \right.
 	\\ \left.\kern-\nulldelimiterspace \times \left[ \frac{\Gamma(j_1+j_2-m+1)}{\Gamma(-j+j_1+j_2+1)} \frac{{_3F_2({m-j},{m_2-j_2},{m+j+1};{m+j_1-j_2+1},{m-j_1-j_2};1 )}}{\Gamma(m+j_1-j_2+1)} \right]^2 \right]\times
 	\\ \times \log \left[ \sum_{m_1=-j_1}^{j_1}\left[ \left[\frac{(j_1+j_2-j)! (2j+1) \Gamma(j+j_1-j_2 +1)   \Gamma(m+j+1)\Gamma(j_2+m_2+1) \Gamma(j_1+m_1+1)}{(j-m)! \Gamma(j-j_1+j_2+1) \Gamma(j+j_1+j_2+2)\Gamma(j_2-m_2+1) \Gamma(j_1-m_1+1)}	
 	\right] \times \right.\right.
 	\\ \left.\kern-\nulldelimiterspace\left.\kern-\nulldelimiterspace \times \left[ \frac{\Gamma(j_1+j_2-m+1)}{\Gamma(-j+j_1+j_2+1)} \frac{{_3F_2({m-j},{m_2-j_2},{m+j+1};{m+j_1-j_2+1},{m-j_1-j_2};1 )}}{\Gamma(m+j_1-j_2+1)} \right]^2 \right]\right] \geq 
 	\\ \geq - \sum_{m_1=-j_1}^{j_1} \sum_{m_2=-j_2}^{j_2}\left[ \left[\frac{(j_1+j_2-j)! (2j+1) \Gamma(j+j_1-j_2 +1)   \Gamma(m+j+1)\Gamma(j_2+m_2+1) \Gamma(j_1+m_1+1)}{(j-m)! \Gamma(j-j_1+j_2+1) \Gamma(j+j_1+j_2+2)\Gamma(j_2-m_2+1) \Gamma(j_1-m_1+1)}	
 	\right] \times \right.
 	\\ \left.\kern-\nulldelimiterspace \times \left[ \frac{\Gamma(j_1+j_2-m+1)}{\Gamma(-j+j_1+j_2+1)} \frac{{_3F_2({m-j},{m_2-j_2},{m+j+1};{m+j_1-j_2+1},{m-j_1-j_2};1 )}}{\Gamma(m+j_1-j_2+1)} \right]^2 \right]\times
 	\\ \times \log \left[\sum_{m_1=-j_1}^{j_1} \sum_{m_2=-j_2}^{j_2}\left[ \left[\frac{(j_1+j_2-j)! (2j+1) \Gamma(j+j_1-j_2 +1)   \Gamma(m+j+1)\Gamma(j_2+m_2+1) \Gamma(j_1+m_1+1)}{(j-m)! \Gamma(j-j_1+j_2+1) \Gamma(j+j_1+j_2+2)\Gamma(j_2-m_2+1) \Gamma(j_1-m_1+1)}	
 	\right] \times \right.\right.
 	\\ \left.\kern-\nulldelimiterspace\left.\kern-\nulldelimiterspace \times \left[ \frac{\Gamma(j_1+j_2-m+1)}{\Gamma(-j+j_1+j_2+1)} \frac{{_3F_2({m-j},{m_2-j_2},{m+j+1};{m+j_1-j_2+1},{m-j_1-j_2};1 )}}{\Gamma(m+j_1-j_2+1)} \right]^2 \right]\right].
 	\end{multline}}
	 
\section{Conclusion}

The main results of this work are the new information-entropic  inequalities for Clebsch-Gordan coefficients of group $SU(2)$. Obtained inequalities and defined in terms of Clebsch-Gordan coefficients mutual information describe the rate of quantum correlations in the system of two spins. We derived the new inequalities for Hahn polynomials and hypergeometric functions $_3F_2$. The suggested approach can be used for obtaining new inequalities for the Clebsch-Gordan coefficients associated with other Lie groups and quantum groups. Applying elaborated in this work approach it is possible to derive new inequalities based on the property of strong subadditivity of entropy of composite system with three subsystems.

\section{Acknowledgments}

VIM acknowledges support from the Tomsk State University Competitiveness Improvement Program.

{\small

\end{document}